\documentstyle[12pt]{article}
\def\tag#1{\space}%
\def\text#1{\space}%
\def\stackunder#1#2{\mathrel{\mathop{#2}\limits_{#1}}}%
\def\bi{\bar{\imath }}
\def\bj{\bar{\jmath }}
\def\bal{\bar{\alpha}}
\def\bbe{\bar{\beta}}

\begin{document}
\begin{titlepage}
\begin{flushright}
Saclay T95/094 \\ MPI-PTh 95-89 \\ September 1995
\end{flushright}
\vskip 1cm
\centerline{{\bf Soft scalar masses in supergravity with horizontal
 $U(1)_X$ gauge symmetry \footnote{Supported in part by the EC grant
Flavourdynamics.}}}
\vskip 24pt
\centerline{{\bf E. Dudas}}
\vskip 8pt
\centerline{CEA, Service de Physique Th\'eorique, CE-Saclay}
\centerline{F-91191 Gif-sur-Yvette Cedex, FRANCE}
\vskip 12pt
\centerline{{\bf S. Pokorski \footnote{On leave from the Institute
of Physics, Warsaw. Supported in part by the Polish Committee for Scientific
Research. }}}
\vskip 8pt
\centerline{Max-Planck Institut f\"ur Physik Werner-Heisenberg-Institut}
\centerline{F\"ohringer Ring 6, D-80805 M\"unchen, GERMANY}
\vskip 12pt
\centerline{{\bf C.A. Savoy}}
\vskip 8pt
\centerline{CEA, Service de Physique Th\'eorique, CE-Saclay}
\centerline{F-91191 Gif-sur-Yvette Cedex, FRANCE}
\vglue 1.2truecm
\begin{abstract}
\vskip 8pt
In supergravity with modular invariance and horizontal $U(1)_X$ gauge symmetry
there is a relation between modular weights and $U(1)_X$ charges. The soft
scalar masses are then strongly correlated with Yukawa matrices. The
implications for FCNC are discussed.
\end{abstract}
\end{titlepage}

\newpage\

1. The observed suppression of flavour changing neutral current (FCNC)
transitions is nicely explained in the Standard Model (SM) by GIM
me\-chanism. At the same time it is a very strong constraint on physics
beyond the SM which, in general, may provide new mechanisms for FCNC
transitions. The supersymmetric extensions of the SM do indeed contain
additional contributions to FCNC transitions from sfermion exchange in loop
diagrams. Such effects are generically suppressed only as $O\left( \frac{M_Z%
}{M_{\widetilde{f}}}\right) $ where $M_{\widetilde{f}}$ is a typical
sfermion mass matrix entry. They can be potentially dangerous for FCNC
transitions if, in the basis in which fermion mass matrices are  diagonal,
the sfermion mass matrices have large flavour off-diagonal entries. The
problem is aggravated by the absence of any reliable theory for calculating
the soft supersymmetry breaking terms. Thus, the solution to the FCNC
problem in supersymmetry remains at the level of speculations which go,
essentially, in the following directions.

Most often explored is the ansatz about universal soft susy breaking terms
at the GUT (or more likely -Planck M$_P)$ scale. This indeed occurs in flavour
blind supersymmetry breaking  scenarios in supergravity \cite{barbieri82}
or
superstrings \cite{guetic91}, \cite{kaplunovski93} (e.g."dilaton  breaking").
At the electroweak scale
we obtain then the flavour dependent effects in the sfermion sector only of
the order of the Cabibbo-Kobayashi-Maskawa mixing, consistently with
observations.  Recently, very interesting progress has been achieved in
studying such ``minimal'' FCNC effects in the framework of GUT's \cite
{barbieri94}.
If we abandon the universality ansatz, we face the problem of explaining in
another way the approximate simultaneous diagonality, in the same basis, of
the fermion and sfermion mass matrices.
For squarks there is the possibility to wash out the flavour
non-universality by large flavour blind renormalization effects from the
scale where supersymmetry is broken to a low-energy scale \cite{brignole1},
\cite{gabbiani89}. Such effects are, however, much weaker
in the slepton  sector and cannot explain the smalness of FCNC unless
flavour dependence of the soft terms at the large scale is for some reason
strictly controlled. It is then conceivable that
the pattern of both types of mass
matrices is simultaneously determined by some symmetries of the lagrangian
\cite{leurer93}, originally introduced in order to understand the hierarchy
of the fermion masses and mixings \cite{froggatt79}.
(Another recent suggestion is the dynamical
allignement \cite{dimopoulos95} of the fermion and sfermion mass matrices at
the electroweak scale.)

Recently \cite{ibanez94}-\cite{nir95}, there is a revival of interest in
explaining the pattern of fermion masses and mixing by postulating a
horizontal $U(1)_X$ gauge symmetry, spontaneously broken at a large scale $%
M.$ The $U(1)_X$ charges are assigned to fermions in such a way that only
a small number of Yukawa interactions is allowed by the symmetry. The
remaining effective Yukawa vertices are generated through non-renormalizable
couplings of the fields which are SM singlets but carry horizontal charge
and spontaneously break $U(1)_X.$ These couplings are suppressed by powers
of a small parameter, $\varepsilon ^{n_i}$,  where the powers $n_i$
depend on the $U(1)_X$ charge assignment.\ In the
context of supersymmetric models with ``stringy'' $U(1)_X$ symmetry \cite
{dine87}-\cite{font88}, this mechanism of fermion mass generation shows an
interesting connection between phenomenologically viable mass pattern and
the Green-Schwarz mechanism of anomaly cancellation, which successfully
predicts the Weinberg angle \cite{ibanez93}.

In this letter we study the predictions for the soft supersymmetry
brea\-king terms which follow from the horizontal $U(1)_X$ symmetry approach
to fermion masses and their implications for FCNC transitions. We work with
the effective supergravity Lagrangian generic for orbifold models of string
compactification, that is we impose on it the spectrum and the symmetries of
the latter. This implies an interesting relation between $U(1)$ charges and
the modular weights of the matter fields, with implications for the
soft supersymmetry breaking terms and FCNC effects.
\vskip .5cm
%%%%%%%%%%%%%%%%%%%%%%%%%%%%%%%%%%%%%%%%%%%%%%%%%%%%%%%%%%%%%%%%%%%%%%%%%%%%%
2. The relevant low energy limit of the superstring models are described by
the $N=1$ supergravity defined by the K\"{a}hler function $K$, the
superpotential $W$ and the gauge kinetic function $f.$ The generic fields
present in the zero-mass string spectrum contain an universal dilaton $%
S $, moduli fields generically denoted by $T_\alpha $ and matter chiral
fields $\phi ^i$ containing the standard model particles. A crucial role in
the following discussion will be played by the target-space modular
symmetries $SL(2,Z)$ \cite{kikkawa84} associated with the moduli
fields $T_\alpha (\alpha =1..m),$ acting as
$ T^\alpha  \rightarrow $$ (a_\alpha T^\alpha -i b_\alpha )$/$(ic_\alpha
T^\alpha +d_\alpha ),$ with $( a_\alpha d_\alpha -b_\alpha c_\alpha )=1 $
 and $ a_\alpha ...d_\alpha \in Z .$  In effective string theories of the
orbifold type, the matter fields $\Phi
^i $ transform under $SL(2,Z)$ as
$\Phi ^i  \rightarrow $$ ( ic_\alpha T^\alpha + d_\alpha )
^{n_i^{(\alpha )}}\Phi ^i $, where the $n_i^{(\alpha )}$ are called the modular
weights of the
fields $\phi ^i$ with respect to the modulus $T^\alpha .$ We define the
{\sl overall modular weight} of the field $\Phi ^i$ by $
n_i = \sum_\alpha n_i^{(\alpha )}.$ These modular transformations, which are
symmetries of the
supergravity theory, can be viewed as a particular type
of K\"{a}hler transformations.

We consider the MSSM model, which we take to be the minimal model obtained
in the low-energy limit of the superstring models, plus the horizontal gauge
group with one singlet $\phi $ of charge $X_\phi .$ We denote the matter
fields by
capitals $\Phi ^i,$ the corresponding $U(1)_X$ charges by small
letters $\varphi _i$ and we define $\varphi _i^{\prime }=-\frac{\varphi _i%
}{X_\phi }.$ The $U(1)_X$ invariant superpotential $W $ and the
 K\"{a}hler potential $K$ read,
\begin{eqnarray}
W  &=&\sum_{ij}\left[ Y_{ij}^U\;\theta \left( q_i^{^{\prime
}}+u_j^{\prime }+h_2^{\prime }\right) \left( \frac \phi M\right)
^{q_i^{\prime }+u_j^{\prime }+h_2^{\prime }}Q^iU^jH_2+\right.
\label{5} \\
&&Y_{ij}^D\;\theta \left( q_i^{\prime }+d_j^{\prime }+h_1^{^{\prime
}}\right) \left( \frac \phi M\right) ^{q_i^{^{\prime
}}+d_j^{\prime }+h_1^{\prime }}Q^iD^iH_1+  \nonumber \\
&&\left. Y_{ij}^E\;\theta \left( \ell _i^{\prime }+e_j^{^{\prime
}}+h_1^{\prime }\right) \left( \frac \phi M\right) ^{\ell _i^{^{\prime
}}+e_j^{\prime }+h_1^{\prime }}L^iE^jH_1\right] ,  \nonumber \\
&&  \nonumber \\
K &=&{K_0}\left( T^\alpha ,\bar{T} ^{\bal} \right) -\ln
\left( S+\bar{S} \right)  + \stackunder{\alpha =1}{\stackrel{p}{\Pi }}t_\alpha
^{n_{\phi }^{(\alpha )}} \bar{\phi } \phi \nonumber \\
&& \ \ + \sum_{\Phi ^i=Q^i,U^i,D^i,L^i,E^i} K_{\bi j}
{\bar{\Phi }}^{\bi } \Phi ^j ,  \nonumber \\
&&  \nonumber \\
K_{\bi j}&=& {\delta}_{\bi j}\stackunder{\alpha =1}{\stackrel{p}{\Pi }}t_\alpha
^{n_j^{(\alpha )}} + {Z}_{\bi j}\left[ \theta \left( \varphi _i^{\prime
}-\varphi
_j^{\prime }\right) \stackunder{\alpha =1}{\stackrel{p}{\Pi }}t_\alpha
^{n_j^{(\alpha )}}\left( \frac{\bar{\phi}}M\right) ^{\varphi _i^{^{\prime
}}-\varphi _j^{\prime }}+\right.  \nonumber \\
&&\left. \theta \left( \varphi _j^{\prime }-\varphi _i^{^{\prime
}}\right) \stackunder{\alpha =1}{\stackrel{p}{\Pi }}t_\alpha ^{n_i^{(\alpha
)}}\left( \frac \phi M\right) ^{\varphi _j^{\prime }-\varphi _i^{^{\prime
}}}\right] + ....  \nonumber
\end{eqnarray}
In (\ref{5}), $M$ is a large mass scale of the order $M_P,t_\alpha $ are the
real part of the $p$ moduli fields $T_\alpha $ and the dots stand for higher
order terms in the fields $\phi$ and $\phi ^i$. Note the flavour
non-diagonal terms in the K\"{a}hler potential, proportional to the
 numbers $Z_{\bi j}$
($Z_{\bi i}= 1$ by a choice of normalization). The coefficients $Z_{\bi j},$
 $Y_{ij}^U,$ $Y_{ij}^D, $ $Y_{ij}^E, $ allowed by the symmetries are
supposed to be naturally of $ O(1) .$

In order to impose the modular symmetries, let us first define
$n_0^{(\alpha )}$ by the modular transformations of the  K\"{a}hler
potential for the modular fields, $K_0 \rightarrow K_0+$
$n_0^{(\alpha )} \ln \left| ic_\alpha T_\alpha +d_\alpha \right| ^2,$
which is a K\"{a}hler transformation.
A typical example, with $n_0^{(\alpha )}=$ $(3/p)$,is
 \begin{eqnarray}
{K}_0 &=&-\frac 3p\sum_{\alpha =1}^p\ln \;t_\alpha .
\label{6} \end{eqnarray}
\noindent Allowing for possible flavour-blind automorphic functions of weight
$ n_W^{(\alpha )}$in the Yukawa couplings,
the modular invariance of $G=K+ln|W|^2$ gives, written explicitly for
the quarks and leptons
\begin{eqnarray}
\theta \left( q_i^{^{\prime
}}+u_j^{\prime }+h_2^{\prime }\right) \left[ \left( q'_i+u'_j+h'_2
\right) n_\phi ^{(\alpha )}+n_{q_i}^{(\alpha )}+n_{u_j}^{(\alpha
)}+n_{h_2}^{(\alpha )}+n_0^{(\alpha )}+n_W^{(\alpha )} \right] &=& 0 ,
\nonumber \\
&&  \nonumber \\
\theta \left( q_k^{^{\prime
}}+d_l^{\prime }+h_1^{\prime }\right) \left[ \left( q'_k+d'_l+h'_1
\right) n_\phi ^{(\alpha )}+n_{q_k}^{(\alpha )}+n_{d_l}^{(\alpha
)}+n_{h_1}^{(\alpha )}+n_0^{(\alpha )}+n_W^{(\alpha )} \right] &=& 0,
\nonumber \\
&&  \nonumber \\
\theta \left( l_m^{^{\prime
}}+e_n^{\prime }+h_1^{\prime }\right) \left[ \left( l'_m+e'_n+h'_1
\right) n_\phi ^{(\alpha )}+n_{l_m}^{(\alpha )}+n_{e_n}^{(\alpha
)}+n_{h_1}^{(\alpha )}+n_0^{(\alpha )}+n_W^{(\alpha )} \right] &=& 0.
\nonumber \\
&&  \label{8}
\end{eqnarray}
Using (\ref{8}), we get in a straightforward way the relation
\begin{eqnarray}
\left( q_i-q_j \right) n_\phi ^{(\alpha )} &=&X_\phi \left(
n_{q_i}^{(\alpha )}-n_{q_j}^{(\alpha )} \right),  \label{9}
\end{eqnarray}
\noindent as a consequence of the existence of the Yukawa couplings
$Y_{ik}^U$ and $Y_{jk}^U$ in the superpotential.
Similar relations are obtained by replacing $q_i$ by $u_i,d_i,l_i,e_i$.
The relations (\ref{9}) gives a surprising connection between the
modular weights and the $U(1)_X$ charges.\ The eq.(\ref{9}) are at the same
time the modular invariance conditions for the existence of the flavour
non-diagonal terms in the K\"{a}hler potential. More exactly, imposing
in the off-diagonal parts of the K\"ahler potential (\ref{5}) the same
weight for the modular transformations of the chiral and the antichiral
parts we get once again eq. (\ref{9}).

If eq.(\ref{9}) are not satisfied, modular invariance of the superpotential
implies zeroes in the Yukawa matrices and
in the off-diagonal entries of the K\"{a}hler metric. These type of zeroes
must be distinguished from the ones given by $U(1)_X$ invariance and
the holomorphicity of the
superpotential $W$ and described by the $\theta $-functions in (\ref{5}). We
could try to construct phenomenologically interesting models in this way, in
the spirit of ref.\cite{ramond93}. An useful rule in this respect is the
following. Consider a $2\times 2$ sub-matrix with three non-zero entries.
Then non-vanishing of the fourth one is automatically consistent with
modular invariance, as a straightforward consequence of eq.(\ref{9}).
This is similar, even though has a very different origin, to the
rule of ref.\cite{casas92}. Using consistently this rule, the only
configurations with zeroes which follow from modular invariance of the
potential (\ref{5}) (up to permutations of lines and
columns which reduce to mere permutations of different generations) are of
the type $\displaystyle \left(
\begin{array}{ccc}
\ast & 0 & 0 \\
0 & \ast & 0 \\
0 & 0 & \ast
\end{array}
\right) $ and $\displaystyle \left(
\begin{array}{ccc}
\ast  & 0 & \ast  \\
0 & \ast  & 0 \\
\ast  & 0 & \ast
\end{array}
\right) .  $
\vskip 5pt

The physical Yukawa couplings, $\hat Y,$ are obtained by the canonical
normalization of
the kinetic terms, which requires the redefinition of the fields
$\hat \Phi ^i = e^i_j \Phi ^j $ where the vielbein $ e^i_j
(t_\alpha , \phi )$
verify
\begin{eqnarray}
K_{\bi j}&=& \delta _{\bar{k} k} \bar{e} ^{\bar{k}}_{\bi} e^k_j =
\stackunder{\alpha =1}{\stackrel{p}{\Pi }}t_\alpha
^{(n_i^{(\alpha )}+n_j^{(\alpha )})/2} \left( {\delta}_{\bi j}+
Z_{\bi j}{ \hat \varepsilon } ^
{| \varphi _i^{\prime }-\varphi _j^{\prime }|} \right) \label{12}
\end{eqnarray}
\noindent where  (\ref{8}) and (\ref{9}) are assumed and the small
parameter ${\hat \varepsilon} = \prod_{\alpha} t_{\alpha}^{n_{
\phi}^{(\alpha)} / 2} {<\phi> \over M}$ serves to estimate the size of
the fermion and scalar mass matrix elements (if, for some $(\bi j),$
(\ref{9}) is not fulfilled, the coefficient vanishes).
The potential effect of these field redefinition  on (\ref{12})
is to remove the eventual zeroes
in the Yukawa matrices (Examples of this type of a phenomenological interest
can be found in \cite{dudas95}).
However, due to the fact that modular symmetry zeroes in Yukawa matrices
imply zeroes
in the corresponding off-diagonal elements of the K\"{a}hler metrics, the
zero textures of the above matrices are preserved after multiplication
by the vielbein. Phenomenologically,
they can acommodate the fermion masses and one
mixing angle, but they cannot explain the whole $V_{CKM}$ matrix. Hence,
for the quarks, the
relations (\ref{9}) must be imposed for {\it all} the indices $(i,j)$  (of
course, zeroes due to the holomorphicity of $W$ can be filled as
in ref.\cite{dudas95}).

The eqs.(\ref{9}) generate, as explained in the next sections, many
equations relating the soft-breaking terms in the low-energy theory.
Interestingly enough, it provides a relation between the hierarchies
within Yukawa coupling matrix elements
and modular weights. Indeed, with canonical normalization for the
fermion fields, one gets,if $n_\phi ^{(\alpha )} \ne 0 $, e.g.,
\begin{eqnarray}
{\hat Y}^U_{ij} &\sim & \hat \varepsilon^{(n_{q_i}^{(\alpha )}-n_{q_3}^{(\alpha
)}
+n_{u_j}^{(\alpha )}-n_{u_3}^{(\alpha )}) /n_{\phi }^{(\alpha )}}
\label{56}
\end{eqnarray}
as well as analogous relations for the matrix elements of $Y^D$ and
$Y^E$ (with an additional factor $Y^D_{33}$ on the r.h.s.).

It would be interesting to compare this approach with that developped in
\cite{binetruy3}, where
the role of the horizontal symmetry is played by the modular symmetries. \vskip
.5cm
%%%%%%%%%%%%%%%%%%%%%%%%%%%%%%%%%%%%%%%%%%%%%%%%%%%%%%%%%%%%%%%%%%%%%%%
3. The spontaneous breaking of local supersymmetry gives rise to a
low-energy global supersymmetric theory together with terms that
explicitly break supersymmetry, but in a soft way. The signal of
supersymmetry breaking is provided by non zero vev's of the auxiliary
components of the chiral superfields $F^a=e^{\frac G2}G^a,$ where $G^a=
K^{a \bar{b} }\partial _{\bar{b} }G.$ We consider only the
case of zero tree level cosmological constant, i.e., we impose $<G^aG_a>=3$
and the order parameter for the supergravity breaking is
provided by the gra\-vi\-tino mass
$m_{3/2}^2=e^G.$ A complete scenario of supersymmetry breaking is still
missing. A pragmatic attitude was taken in \cite{brignole1}, where a
parametrization of the supersymmetry breaking was proposed,
quite independent of its
specific me\-cha\-nism. The fields which participate at the
supergravity breaking were assumed to be the moduli $T^\alpha $
and the dilaton $S.$ The parametrization is
\begin{eqnarray}
G^\beta &=&\sqrt{3} \Theta _\beta  t_\beta  ,  \label{13} \\
  G ^\beta G_\beta &=&3\cos ^2 \theta  ,  \nonumber \\
G^S G_S &=&3\sin ^2\theta .  \nonumber \end{eqnarray}
\noindent The angle $\theta $ and the $\Theta _\alpha $ parametrize the
direction of the goldstino in the $T_\alpha ,S$ space. The normalization
of the $\Theta _\alpha $ is fixed by (\ref{13}).
In the presence of the $U(1)_X$ symmetry spontaneously broken
close to the Planck scale there is an additional
contribution to supersymmetry breaking with $<G_\phi G^\phi> =$
${\hat \varepsilon}^2M^2 / {M_P^2}.$

The soft terms are computed from the usual expressions of supergravity, but
with the flavour non-diagonal K\"{a}hler potential, eq.(\ref{5}). It
is worth noticing that only the lowest power of ${\hat \varepsilon}$ or
$\phi $ have been defined in (\ref{5}) and the calculations are to
remain consistent with this approximation. It goes without saying
that the predictions herebelow have been derived to the lowest power
of ${\hat \varepsilon}$. Since the soft parameters
are relevant for low-energy phenomenology, it is more appropriate
to express them {\sl after the field redefinition} that brings the kinetic
terms to their canonical forms as consistently done herebelow.

Let us first consider the soft scalar masses that, up to a term
proportional to the square of the fermion mass matrices, have the
standard supergravity expression
\begin{eqnarray}
\widetilde{m}^2_{\bi j} = \left( \delta _{\bi j}- G^{\alpha}
{\hat R}_{\bi j \bbe \alpha}
 G^{ \bbe } \right) m_{3/2}^2+g\varphi_i\delta _{\bi j}<D> , \label{150} \\
&&  \nonumber
\end{eqnarray}
where ${\hat R}_{\bi j \bbe \alpha}$ is the Riemann tensor of the K\"ahler
space (with fermion indices for canonical fields) and $<D>$ stands for the
contribution from the D-term of the $U(1)_X$ gauge group, $\displaystyle
D = g\left( X_\phi \left| {\hat \phi }\right| ^2+\right.$ $\displaystyle
\left. \varphi _i\left| {\hat \Phi }_i\right|
^2+\xi \right) $ where $ \xi =
\left( {M_p^2} \over {192 \pi^2}\right) TrX  $ \cite{dine87}.
In the simple case with
only one singlet field $\phi $, the minimization of the potential gives
\begin{eqnarray}
d&=& gX_\phi <D>= -\widetilde{m}^2_\phi=-\left( 1+3 n_{\phi }^{(\alpha )}\Theta
_\alpha ^2 \right)m_{3/2}^2 \ . \
\label{15}
\end{eqnarray}
In this one-singlet model, the $U(1)_X$ symmetry is broken by the Fayet-
Iliopoulos term $\xi $, and this, together with the (assumed)
supersymmetry breaking along the $S$ and $T^\alpha $ components,
induce a component along the $\phi $ direction, with
$<G_\phi G^\phi> = {\hat \varepsilon}^2.$

{}From (\ref{12}), (\ref{13}) and (\ref{150}), one obtains the expression
for the soft scalar mass matrices as follows
\begin{eqnarray}
{1 \over {m_{3/2}^2}} \widetilde{m}_{i\bj }^2 =\left( 1+
3 n_i^{(\alpha )}\Theta_\alpha ^2
-\varphi _i^{\prime } d\right) \delta _{ij}+
3\left| \varphi _i^{\prime }-\varphi_j^{\prime }\right|
Z_{i\bj }n_\phi ^{(\alpha )} \Theta _\alpha ^2
{\hat \varepsilon}^ {\left| \varphi _i^{\prime }
-\varphi _j^{\prime }\right| } \ .   \label{16}
\end{eqnarray}
Remarkably enough, the contribution from supersymmetry breaking along
the $\phi $ direction to (\ref{16}) vanishes.

On account of the non-diagonal form of the K\"{a}hler potential, the
scalar mass matrices are not diagonal as in the usual computations in the
literature.There are two important aspects of this result.
As was discussed in \cite{leurer93}, the horizontal $U(1)_X$ symmetry has
the virtue of  suppresing the flavour off-diagonal terms in the sfermion mass
matrices in a way correlated with the Yukawa matrices. Moreover, in our
model which combines the $U(1)_X$ symmetry with modular invariance the
coefficients of all the entries, including the diagonal ones, are
very constrained and $U(1)_X$ charge dependent.

The soft terms (\ref{16})  depend on the
parameters $n_i^{(\alpha )},\theta ,\Theta _\alpha ,d,{\rm \ etc.}$
In spite of that, one obtains from (\ref{9}) a striking
result for the squark and slepton {\sl mass
differences }. Indeed, by inserting (\ref{15}) and (\ref{9})
into (\ref{16}) one obtains for $\Phi ^i=Q^i,U^i,D^i,L^i,E^i,$
the predictions:
\begin{eqnarray}
\widetilde{m}_{i\bi }^2-\widetilde{m}_{j\bj }^2 &=&(\varphi _i^{\prime }-
\varphi _j^{\prime })m_{3/2}^2 \ . \
\label{18}
\end{eqnarray}
The simple form of this result is due to a
conspiracy between moduli field and D-term contributions in this
one-singlet model.

Also, combining (\ref{8}), (\ref{9}) and (\ref{16}) and introducing the
tree-level gaugino masses $M={\sqrt 3} \sin \theta m_{3/2}$, we obtain
the relations (assuming $n_W^{(\alpha)}=0$)
\begin{eqnarray}
\widetilde{m}_{q_i }^2+\widetilde{m}_{u_j }^2 + \widetilde{m}_{h_2}^2 &=&M^2
+ (q_i^{\prime }+
u_j^{\prime }+h_2^{\prime })m_{3/2}^2 \ . \
\label{180}
\end{eqnarray}
Similar relations are obtained for d-type squarks and sleptons.

There is another relation for soft masses which
follows from the phenomenological approximate equation,
${\rm det}{\hat Y}^D=
{\rm det}{\hat Y}^L.$  The latter translates into a relation
for the $U(1)_X$ charges, \noindent $\stackunder{i}{\sum }\left( q_i+d_i\right)
=$ $\stackunder{i}{\sum }\left( \ell _i+e_i\right) $.
 By subtracting the second
and the third equations in (\ref{8}) one gets the relation
$\displaystyle \sum_i\left( n_{q_i}^{(\alpha )}+n_{d_i}^{(\alpha )}\right) $
=$\displaystyle \sum_i\left(
n_{\ell _i}^{(\alpha )}+n_{e_i}^{(\alpha )}\right) .$ Hence, from
(\ref{16}) one obtains the mass sum rule,
\begin{eqnarray}
\sum_i\left( \widetilde{m}_{\ell _i}^2+\widetilde{m}_{e_i}^2\right)
&=&\sum_i\left( \widetilde{m}_{q_i}^2+\widetilde{m}_{d_i}^2\right) .
\label{21}
\end{eqnarray}
All these equations for scalar masses are to be understood at
energies of the order $M_P,$
and lead to low energy relations after renormalization.

The non-diagonal terms in $K$ and  $W$  affect the trilinear
soft terms $V_{ijk}$, too. The general expression for the trilinear
terms corresponding to fields with $<G^i>=<G_i>=0$ is \cite{zkf},
\begin{eqnarray}
V_{ijk} = \left[ (G^{\alpha} D_{\alpha} + 3) {{W_{ijk}}\over{W}} \right]
m_{3/2}^2 \ ,
\label{160}
\end{eqnarray}
where D stands for the covariant derivative in the K\"{a}hler manifold.
Once again we work with canonical normalization of the scalar fields.
With this convention and in the leading order of the small parameter
$\hat \varepsilon$ the connections in the covariant derivatives in
(\ref{160}) take the simple form,
\begin{eqnarray}
G^{\alpha} \Gamma_{\alpha i}^j = n_i^{(\alpha)} \Theta_{\alpha} \delta_i^j
+ {1 \over 2} \left| \varphi _i^{\prime }-\varphi _j^{\prime }\right|
n_{\phi}^{(\alpha)} \Theta_{\alpha}
{\hat \varepsilon}^
{\left| \varphi _i^{\prime }-\varphi _j^{\prime }\right| } Z_{i\bj } \ .
\label{161}
\end{eqnarray}
The final result for the triscalar coefficient $V_{ia}^U$,
for example, reads
\newpage
\begin{eqnarray}
&& {1 \over m_{3/2}} {\hat V}_{ia}^U =
\sqrt{3} \left[ -\sin \theta +\left( q_i^{\prime } +u_a^{\prime } + h_2^{\prime
}\right) n_\phi^{(\alpha)}\Theta _\alpha \right]{\hat Y} _{ia}^U \label{17}  \\
&&\ \ \ \ \ \ \ \ -\sqrt{3}\left( t_\alpha \frac{\partial
\ln{{\hat Y}_{ia}^U}}
{\partial T_\alpha }
+t_\alpha \partial _{\alpha } K_0 -n_W^{(\alpha)}-n_0^{(\alpha)}\right)
\Theta _\alpha {\hat Y} _{ia}^U+\nonumber \\
&&\ \ \ \ {{\sqrt{3}} \over 2} n_\phi^{(\alpha)} \Theta_{\alpha}\left(
\sum_j | q_i^{\prime }-q_j^{\prime } | Z_{i\bj } {\hat Y}_{ja} {\hat
\varepsilon}^{
|q_i^{\prime }-q_j^{\prime }|} + \sum_b | u_b^{\prime }-u_a^{\prime }|
Z_{a\bar{b} } {\hat Y}_{ib} {\hat
\varepsilon}^{|u_b^{\prime }-u_a^{\prime }|} \right) \nonumber \\
&&\ \ \ \ \ \ \ \ -(q_i^{\prime } +u_a^{\prime } + h_2^{\prime })
{\hat Y} _{ia}^U -(q_i^{\prime } - q_j^{\prime })\theta(q_i^{\prime } -
q_j^{\prime })
Z_{i\bj } {\hat Y}_{ja} {\hat \varepsilon}^{|q_i^{\prime }-q_j^{\prime
}|}\nonumber \\
&&\ \ \ \ \ \ \ \ -(u_a^{\prime } - u_b^{\prime })
\theta(u_a^{\prime } - u_b^{\prime })
Z_{a\bar{b} } {\hat Y}_{ib} {\hat \varepsilon}^{|u_a^{\prime }-u_b^{\prime }|}
\nonumber
\end{eqnarray}
and similar expressions hold for $V^D$ and $V^L$ with obvious replacements.
Notice that the matrices $\hat Y$ have hierarchical entries as expressed
by (\ref{56}). The last terms in (\ref{17}) come from $G^{\phi }D_{\phi },$
namely, from supersymmetry breaking along the $\phi $ direction.
\vskip 0.5cm
%%%%%%%%%%%%%%%%%%%%%%%%%%%%%%%%%%%%%%%%%%%%%%%%%%%%%%%%%%%%%%%%%%%%%%%%%%%%%
4. We now turn to the computation of the FCNC effects in our model. The
fermion masses and the Cabibbo-Kobayashi-Maskawa matrix are obtained by
diagonalization of the mass matrices
$U_L\stackrel{\wedge }{m}^U\stackrel{\wedge }{U}_R^{+} ={\rm diag}%
(m_u,m_c,m_t)$ and the rotation matrices $D_L,\ D_R,\ L_L,\ L_R,$ are defined
analogously.
 The sfermion masses are given by $6\times 6$ matrices $(\widetilde{%
M}^F)^2,\ F=U,D,L,$ which can be divided into $3\times 3$ sub-matrices
\begin{eqnarray}
\left( \widetilde{M}^F\right) ^2 &=&\left(
\begin{array}{cc}
\left( \widetilde{M}^F\right) _{LL}^2 & \left( \widetilde{M}^F\right) _{LR}^2
\\
&  \\
\left( \widetilde{M}^F\right) _{LR}^{2^{+}} & \left( \widetilde{M}^F\right)
_{RR}^2
\end{array}
\right) \;.  \label{23}
\end{eqnarray}

The relevant quantities for the FCNC processes are sfermion off-diagonal
mass matrix elements in the basis in which fermion masses are diagonal, i.e.
\begin{eqnarray}
\left( \delta _{MN}^F\right) _{ij} &=&\frac 1{\widetilde{m}^2}\left[
F_M\left( \widetilde{M}^F\right) _{MN}^2F_N^{+}\right] _{ij}\;,  \label{24}
\end{eqnarray}
\noindent where $M,N=L,R$, $F$ denotes the rotation matrices for fermion
$F$ and
$\widetilde{m}^2$ is an averaged sfermion mass squared.
There exists experimental bounds for the quantities $\delta$ \cite{gabbiani89}.
Typically,
$\left( \delta _{MM}^{d,u}\right) _{ij} \leq \cal{O}\left(
10^{-1}\left( {\tilde m \over 1 TeV}\right)\right) $
with, however, much weaker constraint for the (2,3) element in the down
sector and no constraints for the (1,3) and (2,3) elements in the up sector.
The bounds on the combinations
\begin{eqnarray}
\delta _{ij}^F &=&\sqrt{\left( \delta _{LL}^F\right) _{ij}\left( \delta
_{RR}^F\right) _{ij}}   \label{25}
\end{eqnarray}
are one order of magnitude stronger (on the analogous elements) with
the strongest bound $\delta _{12}^d \leq
 8 \times 10^{-3} \left( {\tilde m / 1 TeV}\right)$.
The best bounds in the lepton sector are
$\left( \delta _{MM}^\ell \right) _{12} \leq 10^{-1}\left( \frac{m_\ell
^{\sim }}{1\;TeV}\right) ^2 $.
We do not discuss the bounds on $\delta_{LR}$. They
constrain the trilinear $V$ soft terms. As was remarked in \cite{leurer93},
the $U(1)_X$ symmetry gives typically $(\delta_{LR}^q)_{ij} \sim {1
\over \tilde m} \sqrt {m_i^q m_j^q}$ leading to very small contributions to
FCNC.

These bounds apply at the electroweak scale. They have to be satisfied by the
soft scalar masses after their renormalization group running from the large
scale at which they are determined by a deeper theory. Eq.(\ref{16})
gives  the sfermion mass matrices at the large (GUT, string) scale, in the
basis defined by the $U(1)_X$ symmetry. The entries in (\ref{16}) are
determined by the $U(1)_X$ charge assignement of sfermions. Here is the
link with the quark Yukawa matrices, determined by the same charge
assignement of the $U(1)_X$ symmetry to fermions.

To go further in the discussion of FCNC one has to specify the model
for Yukawa matrices. Let us consider models with one singlet field.
The acceptable quark Yukawa matrices and the $U(1)_X$ charge
assignements are listed in ref. \cite{dudas95} (they
include the original proposal of Froggatt-Nielsen). To illustrate
the flavour changing problem in these models, we explicitly
calculate $\delta _{12}$'s, for which the constraints are strongest.
Since all those solutions satisfy the relations
$(q_1^{\prime }-q_2^{\prime }) \geq 0$,$(d_1^{\prime }-d_2^{\prime }) \geq
0$,$(u_1^{\prime }-u_2^{\prime }) \geq 0$, one derives the following
results:
\begin{eqnarray}
\left( \delta _{LL}^{u,d} \right) _{12} &=& ( q'_1-q'_2)
(F_{12}^q)   \epsilon^{q'_1-q'_2}  \ , \label{29} \\
\left( \delta _{RR}^u\right) _{12} &=& ( u'_1-u'_2)
(F_{12}^u)  \epsilon^{u'_1-u'_2}  \ , \nonumber \\
\left( \delta _{RR}^d\right)_ {12} &=& ( d'_1-d'_2)
(F_{12}^d)  \epsilon^{ d'_1-d'_2} \ , \nonumber
\end{eqnarray}
where  the $F's$ are obtained from their values at the large scale $M,$
\begin{eqnarray}
F_{ij}^u (M_X) &=& \frac{O(m_{3/2}^2) }{ \left(
\widetilde{m}_{i}^2+\widetilde{m}_{j}^2
\right) } \ .
\label{30}
\end{eqnarray}
through the renormalization group running down to the electroweak scale.
Notice that $\epsilon^{q'_1-q'_2 + u'_1-u'_2 } \sim {m_u
\over m_c}$ and $\epsilon^{q'_1-q'_2 +d'_1-d'_2} \sim {m_d
\over m_s}$ and, consequently, $\left( \delta _{12}^{u} \right)^2$ and
$\left( \delta _{12}^{d} \right)^2$ can be reexpress directly in terms
of the quark masses. For example,
\begin{eqnarray}
\left( \delta _{12}^{d} \right)^2 &=& ( q'_1-q'_2) (d'_1-d'_2)
(F_{12}^q) (F_{12}^d)  {m_d \over m_s} . \label{300}
\end{eqnarray}

The final leading order result is a sum of two terms,
with the same powers of the small parameter $\epsilon$ and proportional to
the charge differences. One is given by the original off-diagonal terms in
(\ref{16}) and the other is proportional to the splittings among the
diagonal entries multiplied by the rotation angles, which are explicitly
given in \cite{dudas95}.
The vanishing of the leading term in the limit of universal
diagonal terms in the sfermion mass matrices reflects an improved allignement
of fermion and sfermion masses due to the $U(1)_X$ symmetry.

In this class of models, the required suppresion of FCNC can be achieved
if the functions $F$ are small enough or if {\em some of the charge
differences in eq.(\ref{300}) vanish}. For instance, $ (\delta _{12}^d)^2 \leq
10^{-4} {\left( {\tilde m \over 1 TeV} \right)}^2$ requires
$(F_{12}^q) (F_{12}^d) \leq 1.6 \ 10^{-3}$ for $\tilde m \sim 1 \ TeV$ and
$(F_{12}^q) (F_{12}^d) \leq 0.6 \ 10^{-4}$ for $\tilde m \sim 0.2 \ TeV$
or one of the equations $(q_1^{\prime }-q_2^{\prime }) = 0$,
$(d_1^{\prime }-d_2^{\prime })= 0$ has to be satisfied.
For small values of the functions $F$ one needs large renormalization
effects in the diagonal squark masses, which requires a sizeable dilaton
supersymmetry breaking .

The second option can also be realized with acceptable quark mass matrices.
Following the classification of ref. \cite{dudas95} of all acceptable
models with one singlet $\phi$ we see that the condition $d_1=d_2$
is satisfied in two models defined by
\begin{eqnarray}
(q_1^{\prime }-q_3^{\prime })=3,(q_2^{\prime }-q_3^{\prime })=2,
(u_1^{\prime }-u_3^{\prime })=5,(u_2^{\prime }-u_3^{\prime })=2,
(d_1^{\prime }-d_3^{\prime })=1 , \label{29} \\
(q_1^{\prime }-q_3^{\prime })=4,(q_2^{\prime }-q_3^{\prime })=3,
(u_1^{\prime }-u_3^{\prime })=4,(u_2^{\prime }-u_3^{\prime })=1,
(d_1^{\prime }-d_3^{\prime })=1  \ . \nonumber
\end{eqnarray}

For these models, $\delta _{12}^d=0$ in the
leading order and the non-leading terms are within the experimental bounds
if $({F^q}_{12}) ({F^d}_{12}) \le .008.$ The limits on
$(\delta _{MM}^{u,d})_{12}$ are also satisfied.
So we conclude that in such models  the FCNC effects are weakened, without
asking for a large dilaton
contribution to supersymmetry breaking.

FCNC effects in the lepton sector are potentially more dangerous as
no strong interaction renormalization effects can wash out the flavour
off-diagonal terms present at the large scale. However, due to our
ignorance on the leptonic mixing angles, the lepton mass models are
much less constrained and many more charge assignements are possible.
Assuming eq.(\ref{9}) to hold for all of the off-diagonal terms in
the lepton sector we get, analogously to (\ref{300})
\begin{eqnarray}
\left( \delta _{12}^\ell \right) ^2 &=&\left( l'_1-l'_2\right) \left(
e'_1-e'_2\right) {F_{12}^l}{F_{12}^e} \left( \frac{m_e}{m_\mu}\right) ^n ,
\label{32}
\end{eqnarray}
where $F^{l,e}$ are similar to $F^{q,d}$ with some obvious
replacements and generically $n=1,2,3$.
The freedom present in the lepton sector makes it relatively easy to
satisfy FCNC constraints, but the analysis must be done model by model.

The relation (\ref{300}) changes for models with more than one singlet and/or
ne\-ga\-tive charge differences.
\vskip 0.5cm
%%%%%%%%%%%%%%%%%%%%%%%%%%%%%%%%%%%%%%%%%%%%%%%%%%%%%%%%%%%%%%%%%%%%%%%%%%%%
5. In this letter, we have discussed effective supergravity models
with horizontal $U(1)_X$ gauge symmetry. It has been shown that the horizontal
symmetry and the modular invariances have to be correlated:
horizontal charges and modular weights
must satisfy eq.(\ref{9}) in order to allow for non-vanishing
Yukawa couplings. The same relations  can be viewed as the conditions for the
existence of flavour non-diagonal terms in the K\"ahler potential.
In turn, the soft supersymmetry breaking terms depend on the $U(1)_X$ charges
and are  correlated with Yukawa matrices.
This results in a predictive framework for the soft  masses.
The FCNC problem for the squarks can be eased
without asking for a large dilaton contribution to supersymmetry breaking.
A systematic phenomenological discussion of different models along these lines
is certainly worthwhile.

\end{document}